\documentclass{aa}
\usepackage{natbib}
\bibpunct{(}{)}{;}{a}{}{,}
\usepackage{graphicx}
\usepackage{txfonts}
\begin{document}
\title{Heavy elements and chemical enrichment in globular clusters\thanks{Based 
    on data collected at the European Southern Observatory with the VLT--UT2, 
    Paranal, Chile (ESO-LP 165.L-0263).}}

\author{G. James\inst{1} \and P. Fran\c{c}ois\inst{1}
  \and P. Bonifacio\inst{2} \and E. Carretta\inst{3} \and R.~G. Gratton\inst{3} 
  \and F. Spite\inst{1}
}

\offprints{G. James, \email{Gael.James@obspm.fr}}

\institute{Observatoire de Paris -- GEPI, 61 Avenue de l'Observatoire, F--75014 Paris, France
  \and INAF -- Osservatorio Astronomico di Trieste, Via G. B. Tiepolo 11, I--34131 Trieste, Italy
  \and INAF -- Osservatorio Astronomico di Padova, Vicolo dell'Osservatorio 5, I--35122 Padova, Italy
}

\date{Received 23 June 2004 / Accepted 05 August 2004}

\abstract{
 High resolution ($R \ga 40\,000$) and high $S/N$ spectra have been acquired with UVES on the 
VLT-Kueyen (Paranal Observatory, ESO Chile) for several main sequence turnoff stars 
($V \sim 17$ mag) and subgiants at the base of the Red Giant Branch ($V \sim 16$ mag) in three 
globular clusters (NGC~6397, NGC~6752 and 47~Tuc/NGC~104) at different metallicities (respectively 
$[{\rm Fe/H}] \simeq -2.0; -1.5; -0.7$). Spectra for a sample of 25 field halo subdwarves have also been taken 
with equal resolution, but higher $S/N$. These data have been used to determine the abundances of several 
neutron-capture elements in these three clusters: strontium, yttrium, barium and europium. This is 
the first abundance determination of these heavy elements for such unevolved stars in these three 
globular clusters. 
These values, together with the [Ba/Eu] and [Sr/Ba] abundance ratios, have been used to test the 
self-enrichment scenario. A comparison is done with field halo stars and other well known Galactic 
globular clusters in which heavy elements have already been measured in the past, at least in bright giants 
($V \ga 11$--12 mag). Our results show clearly that globular clusters have been uniformly 
enriched by $r$-- and $s$--process syntheses, and that most of them seem to follow exactly the same abundance 
patterns as field halo stars, which discards the ``classical'' self-enrichment scenario 
for the origin of metallicities and heavy elements in globular clusters.

  \keywords{Stars: abundances -- Galaxy: globular clusters: 47~Tuc / NGC~104, NGC~6397, NGC~6752}
}

\maketitle
\section{Introduction}

Since almost every evolutionary phase of the stellar life can be found in globular clusters, these 
very old stellar systems are often considered as ideal natural laboratories for testing different theories 
of stellar evolution. It has therefore always been very interesting to compare 
globular cluster stars to other stars that can be observed in the Galaxy, such as field halo, bulge or disk 
stars, and that could present the same kind of chemical history.

Most globular clusters are very homogeneous in the abundances of Fe-peak elements, 
and we could expect them to show abundance patterns quite similar to those of field 
stars at the same metallicities. But in the 1970s, several spectroscopic observations of light metals 
revealed star-to-star inhomogeneities that had not been observed in field stars (\citealt{Osborn1971}; 
and later, \citealt{Norris1981}, who describe a bi-modality in the distribution of CN and CH band strenghts 
in NGC~6752). The first high-resolution observations of cluster stars in the late '70s and the following 
studies of the late '80s and '90s showed variations in the abundances of other (light) elements 
(O, Na, Mg, Al...) in the vast majority of the observed clusters, leading for example to the discovery of 
the O--Na and Mg--Al anticorrelations, and to the first assumptions on the ``primordial'' or ``evolutionary'' 
origins of these so-called ``abundance anomalies'' (e.g. see the first reviews of \citealt{Gra1993}, and 
\citealt{Kraft1994}).

Most of the heavier elements ($Z > 30$) are produced either by slow or rapid neutron-capture reactions (the 
so-called $s$-- and $r$--processes), where both processes occur in different physical conditions and are thus 
likely to happen in different astrophysical sites. Basically, it is commonly assumed that the $r$--process 
elements and a fraction of the light $s$--elements (\textit{weak s}--process) are probably only produced 
at the end of life of high-mass stars, while He-burning in low- and intermediate-mass stars is responsible 
for the production of the other $s$--elements (\textit{main s}--process). But it seems possible that 
other sites or at least different processes could also be involved in the production of the heavy elements 
\citep{Truran2002,Cowan2004,Travaglio2004}. The first spectroscopic observations of neutron-capture elements 
in globular clusters at intermediate- and high-resolution were made in the late '70s, and in the particular 
case of $\omega$~Cen they showed a large dispersion and an overabundance of the $s$--process elements Sr and Ba 
\citep{Dickens1976,Mallia1977}. 
In comparison, previous studies of a few individual field stars had shown an underabundance of the barium to iron 
ratio at $[{\rm Fe/H}] \la -1.5$ \citep{Wallerstein1963,Spite1978}. More recent works have established 
that the barium abundance decreases at low metallicities\footnote{We use the usual spectroscopic notations where $[{\rm A/B}] \equiv {\log}_{10} (N_{\rm A}/N_{\rm B}) - {\log}_{10} {(N_{\rm A}/N_{\rm B})}_{\sun}$, and $\log \epsilon ({\rm A}) \equiv {\log}_{10} (N_{\rm A}/N_{\rm H}) + 12.0$, for two elements A and B. We assume that the term ``metallicity'' is equivalent to the stellar $[{\rm Fe/H}]$ value.}, and becomes subsolar at least in field stars with 
$[{\rm Fe/H}] \la -2.5$ \citep[e.g.][ and references therein]{Gra1994,McW1998,Burris2000,Fulb2002}.

The systematic comparison of heavy element abundance patterns in globular clusters and in field stars 
at different metallicities requires high-resolution spectroscopy and high signal-to-noise observations 
because the transitions of $n$--capture elements are quite hard to detect in cluster stars. In fact, until 
very recently, the size of the largest available telescopes (3--4 m) did not allow easy spectroscopic 
observations at high-resolution and with high $S/N$ ratios for stars much fainter than $V \sim 11$--12 mag. 
In the case of globular clusters, this means that until now it was almost impossible to work with stars others 
than bright giants at the top of the cluster's color-magnitude diagram. And considering that until very recently 
almost no spectrograph had a large UV-visible wavelength coverage, the spectra of globular cluster stars were 
most often obtained only for the yellow-red region, where cool giants emit the most, and where most 
of the transitions necessary to study the different ``abundance anomalies'' of the light metals can be found. 
But most of the $n$--capture elements transitions occur in the violet-blue region, where only 
little flux is available, and which is extremely blended in these cool stars, making such an analysis very difficult 
(the $n$--capture lines are either too strong, or too blended), so that most of the few measurements of heavy elements 
abundances that have been made until now in globular clusters were not very reliable and almost always limited 
to barium and europium, which are the only heavy elements that show a few detectable absorption lines in the red. 
Precise heavy elements abundances can be obtained from weaker and less blended lines. As globular clusters often present 
almost every evolutionary phase of the stellar life, this is possible by going deeper in their color-magnitude 
diagram and searching for scarcely evolved stars (e.g. main-sequence and turnoff stars, subgiants...), 
which are fainter and hotter than bright giants. This can now be achieved with the recent 8--10 m--class 
telescopes (VLT, Subaru, Keck...). 

Testing the chemical evolution of globular clusters has recently made a step forward with the first high-resolution 
and high $S/N$ spectroscopic analyses of globular cluster stars in different evolutionary sequences and at different 
metallicities. The ESO-Large Programme 165.L-0263 (PI: R.~G. Gratton) was dedicated to this purpose. Two groups of 
stars (near the turnoff, and at the base of the RGB, with magnitudes $15 \la V \la 18$) were selected in three 
globular clusters at different metallicities (NGC~6397, NGC~6752 and 47~Tuc, respectively at 
$[{\rm Fe/H}] \simeq -2.0; -1.5; -0.7$). One of the major results concerning the abundance anomalies in globular 
clusters is that the presence of the O--Na anticorrelation has been established also in scarcely evolved stars 
\citep{Gra2001,Carr2004}, showing that these ``anomalies'' are independent of the evolutionary 
phase. This was confirmed by other groups for other clusters \citep[][ and references therein]{Ram2003}. 
These and other results of the recent analyses of scarcely evolved stars in globular clusters are summarized 
in \citet{Gra2004}.

These recent results have been interpreted in the light of different self-pollution scenarios 
\citep{Ventura2001,Dantona2002}, and an emerging hypothesis is that we are observing a second generation of stars, 
formed in a very short period ($\sim${10}$^8$ yrs) from material processed in previous intermediate-mass stars 
(the original cluster population, with the same chemical composition as field stars at the same metallicity) during 
their AGB phase, when the cluster was $\sim${10}$^8$ years old. Such a scenario could explain the abundance 
anomalies observed in globular clusters, but there are still questions about the mass range of these first cluster 
stars. Knowing that no (or very little) variations are seen in the abundances of the $n$--capture elements, in 
particular that no correlations have been found between $n$--capture elements and light metals (Na, Mg...) that show 
abundance anomalies \citep{Armos1994,James2004}, might indicate a restricted mass range of the hypothetical first 
generation of AGB stars. But up to now no model has been developped for the synthesis of heavy elements in globular 
clusters, and the models of self-pollution do not explain the origin of the cluster metallicities. 

\citet{Truran1988} suggested a test for the self-enrichment scenario, which is one of the scenarios that could explain 
the metallicities and the abundances of heavy elements in the clusters: if self-enrichment occured in globular 
clusters, even the most metal-rich clusters would show both high $[\alpha / {\rm Fe}]$ ratios and $r$--process 
dominated heavy element patterns which characterize massive star ejecta as is seen in the most metal-poor stars. 
The aim of this paper, which continues the previous work of \defcitealias{James2004}{Paper~I}\citet[][ hereafter 
referred to as \citetalias{James2004}]{James2004} in the framework of the ESO-LP 165.L-0263, is to determine the ratio 
of the $r$--process to the $s$--process elements in scarcely evolved stars belonging to globular clusters at different 
metallicities in order to test the models of self-enrichment, and to possibly give a clue on the problem of the mass 
range of the first AGBs if a self-pollution scenario is assumed. 
The UVES high-resolution spectrograph at the VLT--UT2 has therefore been extremely interesting because it allowed 
the observation of many scarcely evolved stars in several globular clusters with a very large wavelength coverage. 
These observations permitted the study of the violet-blue region of the spectra, where most of the $n$--capture 
elements transitions can be found. In this work we present new abundance ratios for several $n$--capture elements 
(Sr, Y, Ba and Eu) in stars near the turnoff and in early subgiants of our three main programme clusters: 
47~Tuc (NGC~104), NGC~6752 and NGC~6397.

\section{Observations}
\label{Observations}

\begin{table}
  \centering
  \caption[]{Atmospheric parameters and $S/N$ ratios for our sample of field halo stars \citep[from][]{Gra2003a}.}
  \label{FieldData}
  \begin{tabular}{lccccc}
    \hline\hline
    \noalign{\smallskip}
Star ID   & $S/N$@ & $S/N$@ & $T_{\rm eff}$ & $\log g$ & $\xi$           \\
        & 4500 \AA & 6500 \AA &    (K)           &           & (km s$^{-1}$)   \\
    \noalign{\smallskip}
    \hline
    \noalign{\smallskip}
CD-35 0360 & 123  & 204 &  5048  & 4.53  & {\it1.00}  \\
HD 010607  &  97  & 187 &  5757  & 4.01  & 1.50       \\
HD 029907  & 138  & 189 &  5351  & 4.57  & {\it1.00}  \\
HD 031128  & 158  & 196 &  5970  & 4.45  & {\it1.00}  \\
HD 108177  & 145  & 170 &  6133  & 4.41  & {\it1.00}  \\
HD 116064  & 110  & 184 &  5964  & 4.32  & {\it1.00}  \\
HD 120559  & 135  & 151 &  5383  & 4.57  & {\it1.00}  \\
HD 121004  & 138  & 194 &  5583  & 4.37  & {\it1.00}  \\
HD 126681  & 103  & 208 &  5574  & 4.55  & {\it1.00}  \\
HD 132475  & 134  & 201 &  5541  & 3.79  & 1.27       \\
HD 134169  & 115  & 184 &  5850  & 3.95  & 1.10       \\
HD 134439  & 118  & 179 &  4996  & 4.65  & {\it1.00}  \\
HD 134440  & 103  & 196 &  4714  & 4.61  & {\it1.00}  \\
HD 140283  & 151  & 229 &  5657  & 3.69  & 1.42       \\
HD 145417  &  93  & 174 &  4869  & 4.62  & {\it1.00}  \\
HD 159482  & 128  & 153 &  5713  & 4.35  & 0.96       \\
BD+05 3640 &  83  & 134 &  5023  & 4.61  & {\it1.00}  \\
HD 166913  & 151  & 190 &  6070  & 4.17  & 1.10       \\
HD 181743  & 154  & 165 &  5968  & 4.40  & {\it1.00}  \\
HD 188510  & 147  & 235 &  5503  & 4.55  & {\it1.00}  \\
HD 189558  & 104  & 195 &  5668  & 3.79  & 1.36       \\
HD 193901  & 111  & 183 &  5779  & 4.54  & 1.08       \\
HD 194598  & 132  & 175 &  6023  & 4.31  & 1.26       \\
HD 204155  & 105  & 171 &  5772  & 4.03  & 0.98       \\
HD 205650  & 124  & 156 &  5810  & 4.50  & {\it1.00}  \\
    \noalign{\smallskip}
    \hline
  \end{tabular}
\end{table}

\begin{table}
  \centering
  \caption[]{Main parameters used for the globular clusters \citep[from][]{Gra2001,Carr2004}.}
  \label{ClusterData}
  \begin{tabular}{lcccc}
    \hline\hline
    \noalign{\smallskip}
Cluster ID   & Type$^{\rm (a)}$ & $T_{\rm eff}$ (K)& $\log g$ & $\xi$ (km s$^{-1}$)  \\
    \noalign{\smallskip}
    \hline
    \noalign{\smallskip}
NGC~104 / 47~Tuc & SG &  5134  & 3.65  & 0.90 \\
                 & TO &  5832  & 4.05  & 1.07 \\
NGC~6752         & SG &  5347  & 3.54  & 1.10 \\
                 & TO &  6226  & 4.28  & 0.70 \\
NGC~6397         & SG &  5478  & 3.42  & 1.32 \\
                 & TO &  6476  & 4.10  & 1.32 \\
    \noalign{\smallskip}
    \hline
  \end{tabular} \\
{\footnotesize $^{\rm (a)}$ SG = subgiants, TO = turnoff stars.}
\end{table}

\begin{figure*}
  \centering
  \includegraphics[angle=-90, width=17cm]{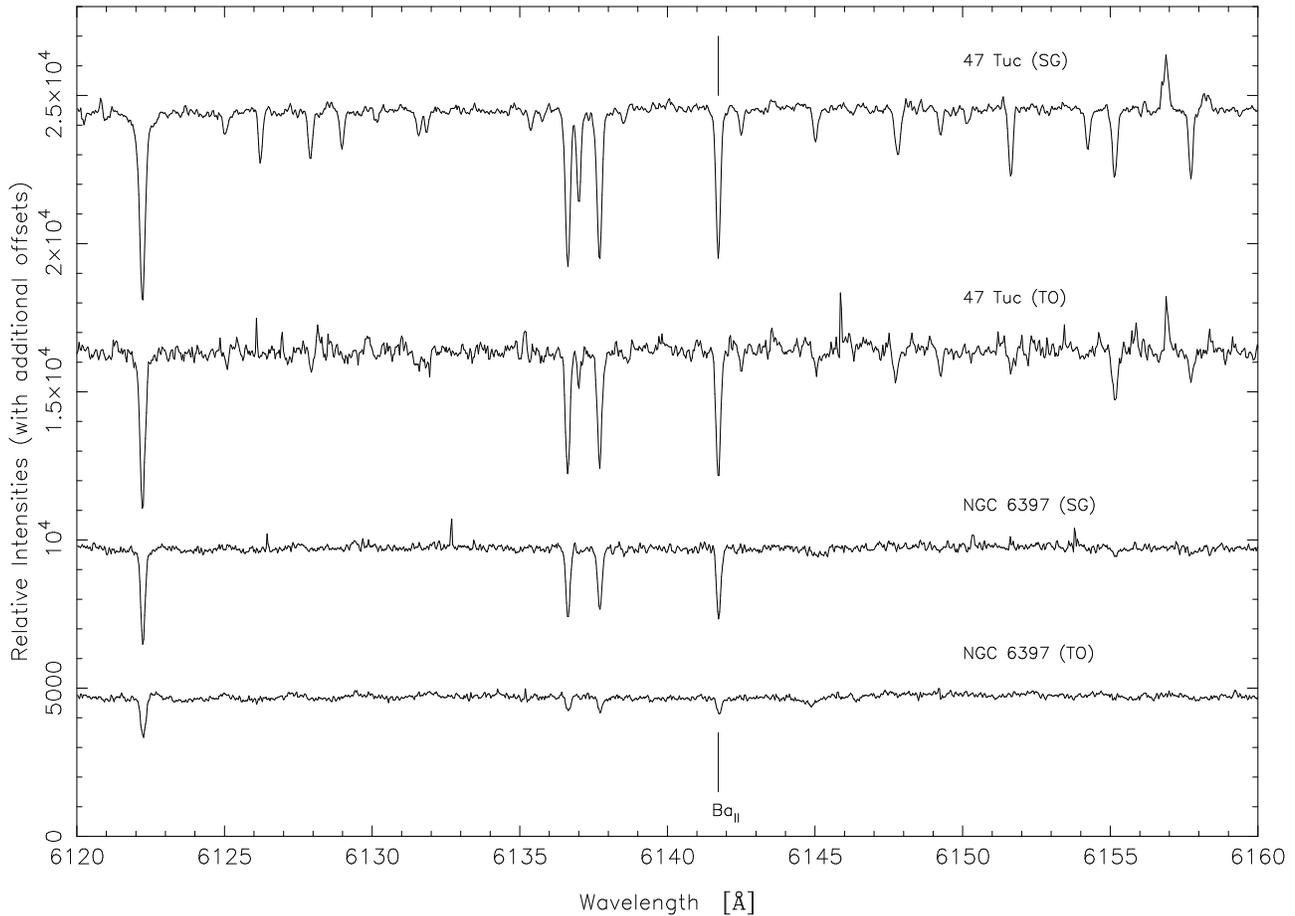}
  \caption{Spectral region near the \ion{Ba}{ii} line at 6141.73 \AA\ for our averaged spectra (UVES, VLT--UT2) in 47~Tuc and NGC~6397. 
    Vertical offsets have been added in order to avoid any overlap and to compare the line intensities 
    between these two clusters, and in each cluster between stars near the TO point and stars at the 
    base of the RGB. Line identification is shown for the \ion{Ba}{ii} line.}
  \label{Balines}
\end{figure*}

\begin{figure*}
  \centering
  \includegraphics[width=5.8cm]{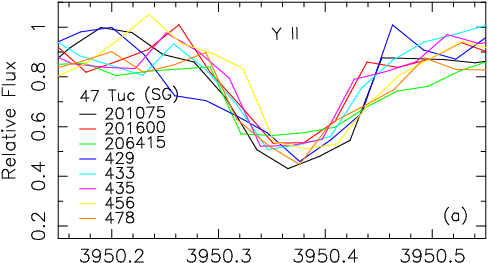}$\ $ \includegraphics[width=5.8cm]{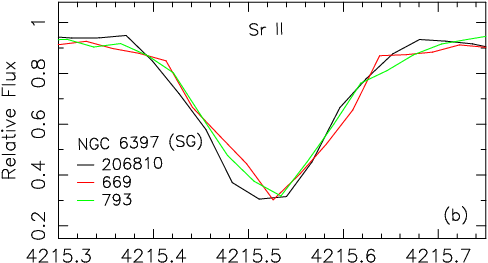}$\ $ \includegraphics[width=5.8cm]{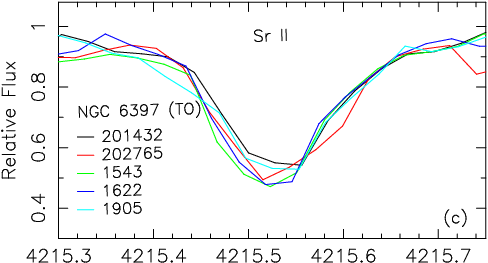}\\
  \vspace{0.2cm}
  \includegraphics[width=5.8cm]{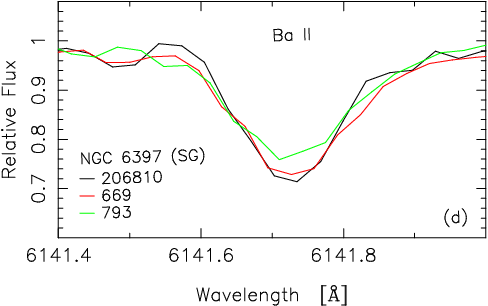}$\ $ \includegraphics[width=5.8cm]{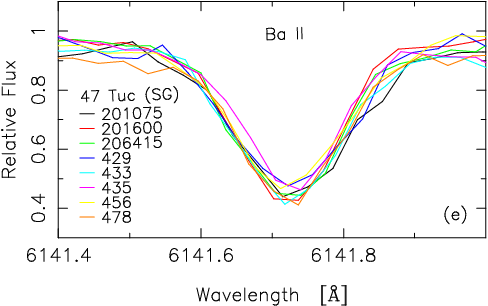}$\ $ \includegraphics[width=5.8cm]{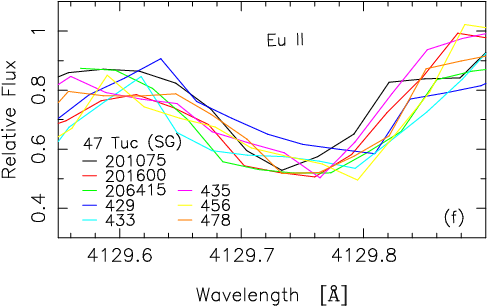}
  \caption{Superimposition of spectra in 47~Tuc and NGC~6397. 
    \textit{Panel (a)}: \ion{Y}{ii} ($\lambda$3950.36) line in the subgiants of 47~Tuc. 
    \textit{Panels (b)} and \textit{(c)}: \ion{Sr}{ii} ($\lambda$4215.52) line in NGC~6397 
    (subgiants and TO stars). 
    \textit{Panels (d)} and \textit{(e)}: \ion{Ba}{ii} ($\lambda$6141.73) line in the 
    subgiants of 47~Tuc and NGC~6397. 
    \textit{Panel (f)}: \ion{Eu}{ii} ($\lambda$4219.70) line in the subgiants of 47~Tuc.
}
  \label{HeavyLines}
\end{figure*}

The observations were carried out in several runs at the Kueyen Telescope (VLT--UT2, ESO) 
with the UVES high-resolution echelle spectrograph between June 2000 and October 2001 in the 
framework of the ESO Large Programme 165.L-0263 (PI: R.~G. Gratton). For NGC~6752 data have been 
obtained for 9 dwarves around the turnoff (TO) and 9 subgiants at the base of the Red Giant Branch (RGB), 
and for NGC~6397 spectra have been collected for 5 TO stars and for 3 subgiants \citep{Gra2001}. 
Finally, for the cluster 47~Tuc (NGC~104), spectra have been acquired for 3 TO stars and 8 subgiants 
as reported in \citet{Carr2004}. A sample of 25 field halo stars was selected to compare the abundance 
patterns relative to our three globular clusters (hereafter ``GCs''). A detailed abundance analysis 
of these field stars can be found in \citet{Gra2003a,Gra2003b}.

Individual cluster stars are listed in previous publications of this Large Programme 
\citep[e.g.][]{Gra2001,Boni2002,Carr2004,James2004}. 
Exposure times are typically of about 1--2 hours for the brightest subgiants in the clusters 
($V \ga 15.5$ mag) and most of the field stars, to about 4--5 hours (split in 3 to 4 exposures) 
for the faintest TO stars ($V \ga 17$ mag). 
The observations have been done with the UVES dichroic beamsplitter \#2, which makes it possible to cover a 
wide spectral range (3500--4700 \AA\ for the blue spectra, and 5700--8700 \AA\ for the red). 
The slit length was always set at 8$\arcsec$. The resolution ($R \equiv \lambda / \Delta \lambda$) 
is always over 40\,000, depending mainly on the slit width which was mostly set at 1$\arcsec$ 
(depending on the seeing). In the blue region near 4500 \AA\ individual $S/N$ ratios range from 
$\sim$15--45 in NGC~6752 and $\sim$20--30 in 47~Tuc to $\sim$50 in NGC~6397. In the red portion 
of the spectra (near 6500 \AA) $S/N$ values spread from $\sim$30--65 in NGC~6752 and $\sim$40 
in 47~Tuc, to $\sim$90--100 in NGC~6397. 

The observed field stars are listed in Table~\ref{FieldData}. Their spectra have been taken using 
the dichroic beamsplitter \#1, and have a resolution of $\sim$50\,000 and $S/N$ ratios around 200 
(see Table~\ref{FieldData}). All our spectra (GCs and field stars) have been 
reduced using the UVES context within ESO--MIDAS\footnote{European Southern Observatory Munich Image 
Data Analysis System, {\tt http://www.eso.org/projects/esomidas/}}.

\section{Abundance analysis}

\subsection{Model atmospheres and stellar parameters}

We have adopted OSMARCS LTE model atmospheres, computed using the grid defined by \citet{Edv1993} 
with the last updated version of the MARCS code of \citet{Gus1975} with improved UV-line 
blanketing \citep[see also][]{Edv1994}. The abundances of \citet{Grev2000} have been used as 
the reference for solar abundances.

For our sample of field stars we have used the atmospheric parameters (effective temperatures 
$T_{\rm eff}$, surface gravities $\log g$, and microturbulent velocities $\xi$) 
published in \citet{Gra2003a}, which we recall in Table~\ref{FieldData}. 
We only had to adjust the microturbulent velocities for some stars (values of $\xi$ written in 
{\it italic} in Table~\ref{FieldData}) to $\xi = 1.0\ {\rm km\ s}^{-1}$ according to comparisons 
with theoretical curves of growth. In addition we have added $S/N$ ratios at 4500 and 6500 \AA\ to show 
the quality of these UVES spectra near regions in which we have searched for heavy element signatures. 

The procedure was different for the GCs. In each one of the three clusters the sample stars 
near the TO point on one hand, and the sample of subgiants on the other hand, are all very close to each 
other in the colour-magnitude (CM) diagram of the corresponding cluster. As a consequence, like in \citet{Gra2001} 
and \citet{Carr2004}, we have used two sets of atmospheric parameters: one for subgiants and one for TO stars. 
Concerning the cluster NGC~6752, we have already described the procedure that was used to check these 
parameters in \citetalias{James2004}. In the same way, we have checked --and then adopted-- the 
parameters published by \citet{Gra2001} for NGC~6397 and by \citet{Carr2004} for 47~Tuc. The only 
difference is that for  this last cluster, unlike in \citet{Carr2004}, we have chosen to use mean parameters 
not only for the TO stars but also for the subgiants. As the effective temperatures determined by 
\citet{Carr2004} for the subgiants are all similar within $\sim$100 K, we have chosen to adopt 
the mean value of $T_{\rm eff} = 5134$ K for these stars. This value was checked 
with the excitation equilibrium method, asking that lines of the same species with different 
excitation potentials should produce the same abundance. 
This method was applied to \ion{Fe}{i} lines with equivalent widths $W_\lambda \le 100\ {\rm m\AA}$.
We have also used a slightly lower value of the gravity ($\log g = 3.65$ instead of 3.84), which 
was adjusted using comparisons to theoretical curves of growth so that 
the ionization equilibrium of iron and titanium were both respected. Table~\ref{ClusterData} 
summarizes the different parameters adopted in each cluster. 
These sets of parameters were then used for each cluster to recompute the metallicity with our OSMARCS code, 
which we finally used as input parameter in our detailed computations.

\subsection{Abundance determinations}

Even with these high-resolution spectra it was not always possible to determine precise abundances 
for several heavy elements in single stars of 47~Tuc due to the low $S/N$ ratios and very 
strong lines at this metallicity ($[{\rm Fe/H}] \sim -0.7\ {\rm dex}$), 
or even to detect weak lines such as those of europium in the faint and metal-poor 
($[{\rm Fe/H}] \sim -2.0\ {\rm dex}$) TO stars of NGC~6397 (see Fig.~\ref{Balines} for a comparison 
of the line intensities in these two clusters) as it had been possible in NGC~6752 
\citepalias[see][]{James2004}.

A careful inspection of the individual spectra revealed that the detected lines of a given 
$n$--capture element (Sr, Y, Ba, or Eu) all have the same intensity in stars of the same sub-sample 
(subgiants or TO stars in each of the two clusters 47~Tuc or NGC~6397). 
Furthermore, when overplotted, the different profiles are almost undistinguishable 
(as is illustrated in Fig.~\ref{HeavyLines}). 
This means that the stars of each sub-sample are not only very similar concerning the atmospheric 
parameters, but  that there is also almost no dispersion in the abundance of these elements in 
the considered sub-samples of the two GCs. As we are using the same atmospheric 
parameters for each single star of a given sub-sample, the computed abundances should be exactly 
the same (within the errors, see Sect.~\ref{Errors}). 
Moreover, adding the spectra increases the $S/N$ per pixel around the lines of the $n$--capture 
elements, making some lines easier to detect and thus providing very accurate mean values for the 
metallicity and also for the abundances of heavy elements in each sub-sample. As a matter of fact, 
this makes it possible to make much better determinations in the noisy spectra of 47~Tuc, and to obtain 
at least upper limits close to detection for the Eu abundance in NGC~6397 
(see Table~\ref{ClusterAbundances}). The $S/N$ ratios in these summed spectra range from 
$\sim$45--55 in the blue and $\sim$70--90 in the red spectra of 47~Tuc, to $\sim$80--90 in the 
blue and $\sim$150--180 in the red spectra of NGC~6397 (see Sect.~\ref{Observations} for a 
comparison with typical individual values).

\begin{table*}
  \centering
  \caption[]{Abundances for our sample of field halo stars.}
  \label{FieldAbundances}
  \begin{tabular}{lccccccccccccc}
    \hline\hline
    \noalign{\smallskip}
Star ID   &  [Fe/H] & [Sr/Fe] & ${{\sigma}_{\rm Sr}}^{\rm (a)}$ & ${N_{\rm Sr}}^{\rm (b)}$ & [Y/Fe] & ${{\sigma}_{\rm Y}}^{\rm (a)}$ & ${N_{\rm Y}}^{\rm (b)}$ & [Ba/Fe] & ${{\sigma}_{\rm Ba}}^{\rm (a)}$ & ${N_{\rm Ba}}^{\rm (b)}$ & [Eu/Fe] & ${{\sigma}_{\rm Eu}}^{\rm (a)}$ & ${N_{\rm Eu}}^{\rm (b)}$ \\
    \noalign{\smallskip}
    \hline
    \noalign{\smallskip}
CD-35 0360 &  --1.17  &  +0.16  &  0.19   &  2  &  +0.19  &  0.03   &  5  &  +0.66  &  0.03  &  3  &  +0.46    &  \ldots  &  1  \\
HD 010607  &  --1.04  &  +0.05  &  0.30   &  2  & --0.21  &  0.11   &  4  &  +0.18  &  0.06  &  3  &  +0.25    &  \ldots  &  1  \\
HD 029907  &  --1.55  & --0.08  &  0.20   &  2  & --0.22  &  0.09   &  5  &  +0.11  &  0.04  &  3  &  +0.66    &  \ldots  &  1  \\
HD 031128  &  --1.55  &  +0.18  &  0.15   &  2  &  +0.04  &  0.06   &  5  &  +0.34  &  0.07  &  3  &  +0.47    &  \ldots  &  1  \\
HD 108177  &  --1.72  &  +0.32  &  0.18   &  2  &  +0.02  &  0.08   &  7  &  +0.14  &  0.07  &  4  & $\le$+0.41 & \ldots  &  1  \\
HD 116064  &  --1.93  &  \ldots &  \ldots &  0  &  +0.83  &  \ldots &  1  &  +0.08  &  0.15  &  4  & $\le$+0.52 & \ldots  &  1  \\
HD 120559  &  --0.89  & --0.05  &  0.13   &  2  & --0.13  &  0.12   &  4  &  +0.15  &  0.12  &  4  &  +0.32    &  \ldots  &  1  \\
HD 121004  &  --0.70  & --0.03  &  0.06   &  2  & --0.07  &  0.08   &  3  &  +0.26  &  0.11  &  4  &  +0.29    &  \ldots  &  1  \\
HD 126681  &  --1.21  &  +0.27  &  0.06   &  2  &  +0.26  &  0.13   &  7  &  +0.60  &  0.11  &  4  &  +0.48    &  \ldots  &  1  \\
HD 132475  &  --1.68  &  +0.30  &  0.11   &  2  &  +0.18  &  0.06   &  7  &  +0.56  &  0.06  &  4  &  +0.35    &  \ldots  &  1  \\
HD 134169  &  --0.82  &  +0.07  &  0.12   &  2  & --0.08  &  0.03   &  3  &  +0.37  &  0.14  &  4  &  +0.24    &  \ldots  &  1  \\
HD 134439  &  --1.38  & --0.26  &  0.23   &  2  & --0.47  &  0.03   &  3  & --0.01  &  0.11  &  4  &  +0.50    &  0.02    &  2  \\
HD 134440  &  --1.27  & --0.55  &  0.39   &  2  & --0.68  &  0.05   &  4  & --0.06  &  0.14  &  4  &  +0.41    &  \ldots  &  1  \\
HD 140283  &  --2.55  & --0.44  &  0.04   &  2  & --0.49  &  0.17   &  3  & --0.90  &  0.04  &  2  &  +0.39    &  \ldots  &  1  \\
HD 145417  &  --1.37  &  \ldots &  \ldots &  0  & --0.05  &  0.12   &  4  &  +0.39  &  0.09  &  3  &  +0.86    &  \ldots  &  1  \\
HD 159482  &  --0.82  &  +0.12  &  \ldots &  1  & --0.04  &  0.09   &  4  &  +0.29  &  0.09  &  4  &  +0.31    &  \ldots  &  1  \\
BD+05 3640 &  --1.10  &  \ldots &  \ldots &  0  &  +0.24  &  0.12   &  4  &  +0.74  &  0.06  &  3  &  +0.39    &  \ldots  &  1  \\
HD 166913  &  --1.64  &  +0.42  &  0.16   &  2  &  +0.22  &  0.08   &  4  &  +0.33  &  0.10  &  4  &  +0.33    &  \ldots  &  1  \\
HD 181743  &  --1.87  &  +0.23  &  0.11   &  2  & --0.01  &  0.05   &  5  &  +0.19  &  0.07  &  2  & $\le$+0.56 & \ldots  &  1  \\
HD 188510  &  --1.55  & --0.04  &  0.17   &  2  & --0.28  &  0.07   &  4  &  +0.25  &  0.12  &  4  &  +0.39    &  \ldots  &  1  \\
HD 189558  &  --1.19  &  +0.28  &  \ldots &  1  &  +0.07  &  0.11   &  6  &  +0.51  &  0.13  &  4  &  +0.33    &  \ldots  &  1  \\
HD 193901  &  --1.14  & --0.01  &  0.22   &  2  & --0.17  &  0.09   &  6  &  +0.36  &  0.16  &  4  &  +0.48    &  \ldots  &  1  \\
HD 194598  &  --1.18  & --0.16  &  0.53   &  2  & --0.19  &  0.09   &  6  &  +0.28  &  0.10  &  4  &  +0.32    &  \ldots  &  1  \\
HD 204155  &  --0.73  &  \ldots &  \ldots &  0  &  +0.07  &  0.08   &  3  &  +0.37  &  0.10  &  4  &  +0.27    &  \ldots  &  1  \\
HD 205650  &  --1.20  &  +0.10  &  0.17   &  2  &  +0.07  &  0.10   &  5  &  +0.37  &  0.14  &  4  &  +0.61    &  0.11    &  2  \\
    \noalign{\smallskip}
    \hline
  \end{tabular} \\
{\footnotesize $^{\rm (a)}$ ${\sigma}_{\rm X}$ = standard deviation around the mean value; $^{\rm (b)}$ $N_{\rm X}$ = number of lines used for the element X.} \\
\end{table*}

\begin{table*}
  \centering
  \caption[]{Mean abundances for the globular clusters NGC~104 (47~Tuc), NGC~6752, and NGC~6397.}
  \label{ClusterAbundances}
  \begin{tabular}{lcccccccccccc}
    \hline\hline
    \noalign{\smallskip}
Cluster ID & Type$^{\rm (a)}$ & ${N_*}^{\rm (b)}$ & [Fe/H] & ${{\sigma}_{\rm Fe}}^{\rm (c)}$ & [Sr/Fe] & ${{\sigma}_{\rm Sr}}^{\rm (c)}$ & [Y/Fe] & ${{\sigma}_{\rm Y}}^{\rm (c)}$ & [Ba/Fe] & ${{\sigma}_{\rm Ba}}^{\rm (c)}$ & [Eu/Fe] & ${{\sigma}_{\rm Eu}}^{\rm (c)}$ \\
    \noalign{\smallskip}
    \hline
    \noalign{\smallskip}
NGC~104 / 47~Tuc   & SG  & 8 & --0.69  &  0.06  &  +0.36  &  0.16  & --0.11  &  0.10   &  +0.35  &  0.12  &  +0.17    &  \ldots  \\ 
                   & TO  & 3 & --0.68  &  0.01  &  +0.28  &  0.14  &  +0.06  &  0.01   &  +0.22  &  0.12  &  +0.11    &  \ldots  \\
    \noalign{\smallskip}
NGC~6397           & SG  & 3 & --2.08  &  0.01  & --0.04  &  0.14  & --0.20  &  0.09   & --0.17  &  0.06  & $\le$+0.41    &  0.09    \\
                   & TO  & 5 & --2.07  &  0.01  & --0.06  &  0.15  & $\le$--0.17 & 0.13 & --0.31 &  0.06  & $\le$+0.46    &  \ldots  \\
    \noalign{\smallskip}
    \hline
    \noalign{\smallskip}
    \multicolumn{12}{c}{Data for NGC~6752 \citep[from][]{James2004}.} \\
    \noalign{\smallskip}
    \hline
    \noalign{\smallskip}
NGC~6752           & SG  & 9 & --1.49  &  0.07  & --0.01  &  0.09   & --0.01  &  0.13   &  +0.25  &  0.08  &  +0.40    &  0.09   \\
                   & TO  & 9 & --1.48  &  0.07  &  +0.12  &  0.19   & --0.03  &  0.11   &  +0.11  &  0.09  &  +0.47    &  0.08   \\
    \noalign{\smallskip}
    \hline
  \end{tabular} \\
{\footnotesize $^{\rm (a)}$ SG = subgiants, TO = turnoff stars; $^{\rm (b)}$ $N_*$ = number of stars in each group; $^{\rm (c)}$ ${\sigma}_{\rm X}$ = standard deviation around the mean value.}
\end{table*}

\subsection{Iron abundances}

The iron abundances for all our stars (field sample and clusters) were deduced from equivalent 
width measurements made with an automatic line fitting procedure based on the algorithms of 
\citet{Char1995}, which perform both line detection and gaussian fits of lines including blends. 
We made a selection of $\sim$30--40 (depending on the quality of the spectra) \ion{Fe}{i} 
lines with $W_\lambda \le 100\ {\rm m\AA}$ and kept all the detected \ion{Fe}{ii} 
lines. 

Table~\ref{FieldAbundances} lists iron abundances (the average of \ion{Fe}{i} and \ion{Fe}{ii}) 
in our field stars, and Table~\ref{ClusterAbundances} gives the mean values in the clusters. 
All these metallicities are fully compatible with previous studies 
\citep[][ and references therein]{Gra2001,Boni2002,Gra2003a,Carr2004}. 
We note that there is almost no dispersion between the metallicities found in the subgiants or 
in the TO stars, especially in NGC~6397, as was emphasized by \citet{Gra2001} and \citet{Boni2002}.

\subsection{Heavy element abundances}

Heavy element abundances were measured in the same way for field and cluster stars. 
Strontium, yttrium, barium and europium lines were detected using the same automatic line fitting 
procedure as for iron lines, but abundances were determined using synthetic spectra: we used 
an updated version of the synthetic spectrum code of \citet{Spite1967}. This allowed us 
to take into account the hyperfine structure (hereafter ``hfs'') of several barium and europium lines 
\citep[e.g.][ and references therein]{McW1998}, and it was also more efficient for the very 
strong and broad lines of strontium where a simple gaussian fit did not always match the line profile. 
The computations were done using LTE model atmospheres but we are aware that heavy elements can be 
affected by non-LTE \citep[e.g.][]{Mash1999,Mash2000,Mash2001}. At intermediate metallicities, 
in most of the cases the corrections to be applied do not exceed $\pm$0.15 dex. The size of 
these corrections depends on the observed line and on the atmospheric parameters, but in the present 
case, as we are especially interested in abundance determinations in GCs, regarding the 
uncertainties in the atmospheric parameters in our clusters combined with the error in the 
$W_\lambda$ measurements, a real change in the abundance patterns of our sample is not expected.

The line parameters for \ion{Sr}{ii} and \ion{Y}{ii} were taken from \citet{Sneden2003}. 
Strontium abundances were obtained from the lines at 4077.71 and 4215.52 \AA, and yttrium abundances 
were deduced from the lines at 3950.36 and 4398.01 \AA\ in the clusters, and for the field stars also 
from the lines at 3549.01, 3600.74, 3611.04, 3774.33, and 3788.70 \AA. 
Barium and europium hfs parameters were taken respectively from \citet{McW1998} and 
\citet{Lawler2001}. The \ion{Ba}{ii} lines at 5853.69, 4554.03, 6141.73 and 6496.91 \AA\ and the 
\ion{Eu}{ii} line at 4129.70 \AA\ could almost always be detected in the whole sample (field and 
cluster stars). Sometimes we also had the \ion{Eu}{ii} line at 4205.05 \AA, but it was almost always 
very strongly blended.

The abundances of these four heavy elements are given in Tables~\ref{FieldAbundances} 
and \ref{ClusterAbundances}. Figure~\ref{GCvsF_all} displays the ratios of $n$--capture element 
abundances to iron as a function of the metallicity. These abundance patterns and the comparison 
between GCs and field halo stars will be discussed in the following sections.

\subsection{Error estimations}
\label{Errors}

\begin{table}
  \centering
  \caption[]{Typical error estimates for field and cluster stars.}
  \label{TabErrors}
  \begin{tabular}{lcccc}
    \hline\hline
    \noalign{\smallskip}
    El.  & $\Delta T_{\rm eff}$ & $\Delta \log g$ & $\Delta \xi$     & $\Delta ({\rm Tot})$\\
         &  +100 K                 & +0.2 dex        & +0.2 km s$^{-1}$ &    \\
    \noalign{\smallskip}
    \hline
    \noalign{\smallskip}
    \multicolumn{5}{c}{Field stars (e.g. HD132475)}\\
    \noalign{\smallskip}
    \hline
    \noalign{\smallskip}
    \ion{Fe}{i}  & +0.09 & --0.01 & --0.03 &  0.10 \\
    \ion{Fe}{ii} & +0.01 &  +0.09 & --0.03 &  0.10 \\
    \noalign{\smallskip}
    \hline
    \noalign{\smallskip}
    \ion{Sr}{ii} & +0.09 & --0.02 & --0.02 &  0.09 \\
    \ion{Y}{ii}  & +0.06 &  +0.07 & --0.10 &  0.14 \\
    \ion{Ba}{ii} & +0.07 &  +0.02 & --0.10 &  0.12 \\
    \ion{Eu}{ii} & +0.05 &  +0.08 & --0.02 &  0.10 \\
    \noalign{\smallskip}
    \hline
    \noalign{\smallskip}
    \multicolumn{5}{c}{47~Tuc -- Subgiants} \\
    \noalign{\smallskip}
    \hline
    \noalign{\smallskip}
    \ion{Fe}{i}  &  +0.10 & --0.04 & --0.09 &  0.14 \\
    \ion{Fe}{ii} & --0.06 &  +0.08 & --0.06 &  0.12 \\
    \noalign{\smallskip}
    \hline
    \noalign{\smallskip}
    \ion{Sr}{ii} &  +0.04 &  +0.01 & --0.03 &  0.05 \\
    \ion{Y}{ii}  &  +0.03 &  +0.07 & --0.12 &  0.14 \\
    \ion{Ba}{ii} &  +0.04 &  +0.01 & --0.10 &  0.11 \\
    \ion{Eu}{ii} &  +0.03 &  +0.07 & --0.09 &  0.12 \\
    \noalign{\smallskip}
    \hline
    \noalign{\smallskip}
    \multicolumn{5}{c}{47~Tuc -- TO stars} \\
    \noalign{\smallskip}
    \hline
    \noalign{\smallskip}
    \ion{Fe}{i}  &  +0.09 & --0.04 & --0.08 &  0.13 \\
    \ion{Fe}{ii} & --0.01 &  +0.06 & --0.07 &  0.09 \\
    \noalign{\smallskip}
    \hline
    \noalign{\smallskip}
    \ion{Sr}{ii} &  +0.06 & --0.01 & --0.02 &  0.06 \\
    \ion{Y}{ii}  &  +0.04 &  +0.07 & --0.10 &  0.13 \\
    \ion{Ba}{ii} &  +0.05 &  +0.03 & --0.11 &  0.12 \\
    \ion{Eu}{ii} &  +0.03 &  +0.08 & --0.05 &  0.10 \\
    \noalign{\smallskip}
    \hline
    \noalign{\smallskip}
    \multicolumn{5}{c}{NGC~6397 -- Subgiants} \\
    \noalign{\smallskip}
    \hline
    \noalign{\smallskip}
    \ion{Fe}{i}  & +0.10 & --0.01 & --0.07 &  0.12 \\
    \ion{Fe}{ii} & +0.02 &  +0.08 & --0.04 &  0.09 \\
    \noalign{\smallskip}
    \hline
    \noalign{\smallskip}
    \ion{Sr}{ii} & +0.10 &  +0.01 & --0.13 &  0.16 \\
    \ion{Y}{ii}  & +0.07 &  +0.08 & --0.08 &  0.13 \\
    \ion{Ba}{ii} & +0.07 &  +0.05 & --0.06 &  0.10 \\
    \ion{Eu}{ii} & +0.04 &  +0.07 & --0.01 &  0.08 \\
    \noalign{\smallskip}
    \hline
    \multicolumn{5}{c}{NGC~6397 -- TO stars} \\
    \noalign{\smallskip}
    \hline
    \noalign{\smallskip}
    \ion{Fe}{i}  & +0.08 & --0.04 & --0.07 &  0.11 \\
    \ion{Fe}{ii} & +0.01 &  +0.06 & --0.03 &  0.07 \\
    \noalign{\smallskip}
    \hline
    \noalign{\smallskip}
    \ion{Sr}{ii} & +0.05 &  +0.03 & --0.18 &  0.19 \\
    \ion{Y}{ii}  & +0.05 &  +0.07 & --0.01 &  0.09 \\
    \ion{Ba}{ii} & +0.05 &  +0.05 & --0.03 &  0.08 \\
    \ion{Eu}{ii} & +0.05 &  +0.06 & --0.03 &  0.08 \\
    \noalign{\smallskip}
    \hline
  \end{tabular}
\end{table}

We assumed that the total error budget was due to random uncertainties in the oscillator strenghts 
($gf$ values) and in the measurement of the equivalent widths (or to fitting uncertainties), 
and to errors in the stellar parameters. When $N \ge 2$ lines of a given element are observed, 
the random uncertainties can be estimated using the standard deviation around the mean abundance 
(see Tables~\ref{FieldAbundances} and \ref{ClusterAbundances}). 
The errors linked to the uncertainties in the stellar atmosphere parameters were estimated 
assuming the following variations: 
$\Delta T_{\rm eff} = \pm 100\ {\rm K}$, $\Delta \log g = \pm 0.2\ {\rm dex}$, and 
$\Delta \xi = \pm 0.2\ \mbox{km s}^{-1}$. 

Table~\ref{TabErrors} summarizes the typical error estimates for field stars and for GC stars. 
Computations of the errors for the other stars of our sample give similar results. 
The total error given in the last column of Table~\ref{TabErrors}, $\Delta ({\rm Tot})$,  
is the quadratic sum of all the individual uncertainties linked to the atmospheric parameters, 
and does not take into acount the uncertainty coming from the measurements of the equivalent 
widths (or the fitting errors). From this table, we can see that the total error $\Delta ({\rm Tot})$ 
due all the uncertainties in the atmospheric parameters almost never exceeds 0.15 dex.

\section{Results}
\label{Results}

\subsection{Abundance patterns}

\begin{figure}
  \centering
  \includegraphics[width=8.5cm]{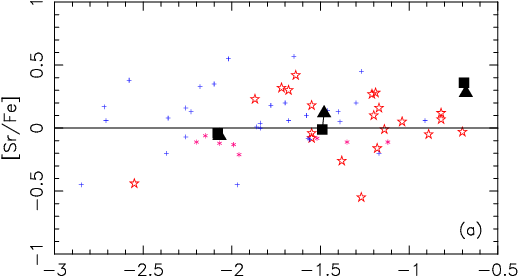}\\ \vspace{0.1cm}
  \includegraphics[width=8.5cm]{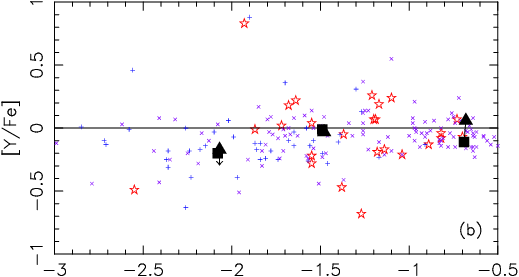}\\ \vspace{0.1cm}
  \includegraphics[width=8.5cm]{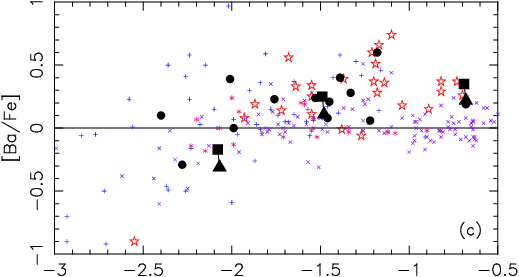}\\ \vspace{0.1cm}
  \includegraphics[width=8.5cm]{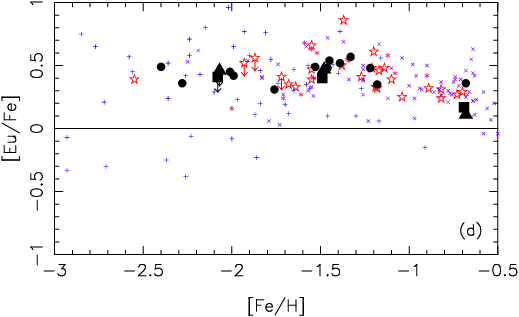}
  \caption{[X/Fe] mean ratios for the $n$--capture elements in Galactic globular clusters compared to field stars. 
    \textit{Filled squares} ($\blacksquare$): subgiants in 47~Tuc, NGC~6752 and NGC~6397. 
    \textit{Filled triangles} ($\blacktriangle$): TO stars in the three clusters (a line joins the two mean 
    points for each globular cluster). 
    \textit{Open stars} ({\small$\star$}): our sample of field halo stars. 
    \textit{Small crosses} ({\tiny$+$}, {\tiny$\times$}, {\small$\ast$}): other field stars taken from the 
    literature \citep{Burris2000,Fulb2000,Mash2001,Mash2003}.
    \textit{Filled circles} ($\bullet$): other Galactic globular clusters (with increasing metallicity: M~15, M~92, 
    NGC~6287, NGC~6293, NGC~6541, M~13, M~79, M~3, NGC~288, NGC~362, M~5, M~4 and M~71); abundances are 
    from \citet{Francois1991}, \citet{Armos1994}, \citet{Shet1996}, \citet{Sneden1997}, 
    \citet{Ivans1999,Ivans2001}, \citet{Shet2000}, \citet{Lee2002}, \citet{Ram2003} and \citet{Sneden2004}.
  }
  \label{GCvsF_all}
\end{figure}

Figure~\ref{GCvsF_all} displays the [$n$--capture/Fe] ratios as a function of the 
metallicity for our sample stars. To increase the field halo star sample, we have added data points 
from \citet{Burris2000}, \citet{Fulb2000}, and the halo stars of \citet{Mash2001} and 
\citet{Mash2003}. We have used the following symbols for Figs.~\ref{GCvsF_all} and 
\ref{Heavy_Ratios}: 
\textit{Filled squares} ($\blacksquare$) indicate the subgiants in our three GCs, 
and \textit{filled triangles} ($\blacktriangle$) indicate the TO stars in the same clusters 
(for a given cluster, the two data points are joined by a line); 
\textit{open stars} ({\small$\star$}) stand for our field stars, and other 
GCs and field stars --from the literature-- are respectively represented by 
\textit{filled circles} ($\bullet$) and \textit{small crosses} ({\tiny$+$}, 
{\tiny$\times$}, {\small$\ast$}); limits are always indicated by an arrow plotted on one 
of the previous symbols. The data for other GCs that we show here for comparison 
have all been obtained from the analyses of bright giant stars near the top of the color-magnitude 
diagrams of the corresponding clusters. Only the analysis of \citet{Ram2003} goes as deep 
as ours and provides data for some heavy elements in M~71 and M~5, and also barium for some stars near 
the turnoff point of these two clusters.

\begin{description}
\item \textit{Strontium (Fig.~\ref{GCvsF_all}a)}: 
      there is some (expected, see \citealt{Ryan1991}) spread in the [Sr/Fe] ratios of our 
      field halo stars, but these ratios still follow the general trend found in other studies 
      \citep[e.g.][ and references therein]{Burris2000,Mash2003}. Strontium abundances 
      could be determined for almost the whole sample, with the exception of four stars --owing to strong 
      blends in the \ion{Sr}{ii} lines-- (see Table~\ref{FieldAbundances}). Most of the strontium 
      to iron ratios are solar or slightly oversolar. 

      The three GCs also follow this trend. 
      NGC~6397 and NGC~6752 have both [Sr/Fe] ratios very close to the solar value: \\
      ${\rm {[Sr/Fe]}_{(SG)}} = -0.04 \pm 0.14$ and ${\rm {[Sr/Fe]}_{(TO)}} = -0.06 \pm 0.15$\\
      for NGC~6397;\\
      ${\rm {[Sr/Fe]}_{(SG)}} = -0.01 \pm 0.09$ and ${\rm {[Sr/Fe]}_{(TO)}} = +0.12 \pm 0.19$\\
      for NGC~6752, and 47~Tuc seems definitively to have a rather oversolar strontium to iron ratio:\\
      ${\rm {[Sr/Fe]}_{(SG)}} = +0.36 \pm 0.16$ and ${\rm {[Sr/Fe]}_{(TO)}} = +0.28 \pm 0.14$.\\ 
      The abundance spreads among the two star types in the clusters (subgiants or dwarves) are 
      comparable, but as we already noticed in \citetalias{James2004}, there seems to be a larger 
      spread for dwarves than for subgiants in NGC~6752. We can add that the abundance spreads are 
      probably due in major part to the difficulty of positioning the continuum level in the fits 
      of the \ion{Sr}{ii} lines in the three clusters. In any case, the standard deviations given 
      here (see also Table~\ref{ClusterAbundances}) are larger than the difference between the 
      ratios in the subgiants and in the dwarves, which we can remark is quite small ($\la$ 0.1 dex).\\

\item \textit{Yttrium (Fig.~\ref{GCvsF_all}b)}: 
      our field halo sample abundance distribution agrees well with previous studies at 
      intermediate metallicity \citep[][ and references therein]{Burris2000,Fulb2002}. 
      The dispersion in the yttrium to iron ratios is comparable to what is found in other 
      studies at this metallicity range, and almost all our field sample stars have [Y/Fe] 
      ratios concentrated around the solar value. 

      We can notice that the GCs follow 
      this trend and also have values very close to the solar ratio. Yttrium abundances were 
      obtained for almost each star type in the three clusters, and we obtained an upper limit 
      for the TO stars in NGC~6397:\\
      ${\rm {[Y/Fe]}_{(SG)}} = -0.20 \pm 0.09$ and ${\rm {[Y/Fe]}_{(TO)}} \le -0.17 \pm 0.13$\\
      for NGC~6397;\\
      ${\rm {[Y/Fe]}_{(SG)}} = -0.01 \pm 0.13$ and ${\rm {[Y/Fe]}_{(TO)}} = -0.03 \pm 0.11$\\
      for NGC~6752;\\
      ${\rm {[Y/Fe]}_{(SG)}} = -0.11 \pm 0.10$ and ${\rm {[Y/Fe]}_{(TO)}} = +0.06 \pm 0.01$\\
      for 47~Tuc. 
      For the two clusters NGC~6752 and NGC~6397, the standard deviations are larger than the 
      dispersion between the ratios in the different stellar types, while in the case of 47~Tuc 
      we note a small offset (0.17 dex) in the [Y/Fe] ratio between the subgiants and the dwarves.\\

\item \textit{Barium (Fig.~\ref{GCvsF_all}c)}: 
      barium abundances have been determined for all the field halo stars. Most of them have solar or 
      oversolar barium to iron ratios, which is in good agreement with previous publications of 
      metal-poor star analyses in this metallicity range \citep{Burris2000,Fulb2002,Mash2003}. 
      The dispersion found among our sample stars for a given metallicity agrees also with 
      previous data. 

      The three GCs follow exactly the same trend. We find:\\
      ${\rm {[Ba/Fe]}_{(SG)}} = -0.17 \pm 0.06$ and ${\rm {[Ba/Fe]}_{(TO)}} = -0.31 \pm 0.06$\\
      for NGC~6397;\\
      ${\rm {[Ba/Fe]}_{(SG)}} = +0.25 \pm 0.08$ and ${\rm {[Ba/Fe]}_{(TO)}} = +0.11 \pm 0.09$\\
      for NGC~6752;\\
      ${\rm {[Ba/Fe]}_{(SG)}} = +0.35 \pm 0.12$ and ${\rm {[Ba/Fe]}_{(TO)}} = +0.22 \pm 0.12$\\
      for 47~Tuc. 
      For the three clusters, we note a small offset between the subgiants and the dwarves 
      (0.14 dex in NGC~6397 and NGC~6752 --see also \citetalias{James2004}--, and 0.13 in 47~Tuc), 
      the latter having always the lower value in a given cluster. 
      From these data points only, we could easily conclude that the [Ba/Fe] ratio in Galactic 
      GCs undergoes a soft decline as the metallicity decreases below --2.0, just 
      as is seen in the metal-poor field stars \citep[][ and references therein]{Gra1994,Fulb2002}, 
      but a quick comparison to previous abundance analyses of giants in Galactic GCs 
      \citep{Armos1994,Shet1996,Sneden1997,Sneden2000,Sneden2004,Ivans1999,Shet2000,Ivans2001,Lee2002,Ram2003}
      shows that the existing dispersion of the [Ba/Fe] ratio becomes larger at low metallicities, 
      so that no real conclusion can yet be drawn. Anyway, these studies give [Ba/Fe] ratios in giants 
      ranging from --0.29 to +0.60 in GCs with metallicities between --2.40 and --0.68, 
      and our analysis is wholly compatible with these values.\\

\item \textit{Europium (Fig.~\ref{GCvsF_all}d)}: 
      europium abundances have been determined for almost the whole sample, and upper limits are 
      given for the only three field halo stars where no reliable measure was possible. The 
      abundance spread is noticeably smaller than for the other $n$--capture elements in the field 
      sample. The field stars all have oversolar europium to iron ratios concentrated around 
      +0.40 dex, and they agree completely with the ratios given by other studies at intermediate 
      and low metallicity \citep[][ and references therein]{Fulb2002,Mash2003}. 

      The GCs follow this trend remarkably well as we find (upper limits have been 
      preferred for NGC~6397):\\
      ${\rm {[Eu/Fe]}_{(SG)}} \le +0.41 \pm 0.09$ and ${\rm {[Eu/Fe]}_{(TO)}} \le +0.46 \pm 0.00$\\
      for NGC~6397;\\
      ${\rm {[Eu/Fe]}_{(SG)}} = +0.40 \pm 0.09$ and ${\rm {[Eu/Fe]}_{(TO)}} = +0.47 \pm 0.08$\\
      for NGC~6752;\\
      ${\rm {[Eu/Fe]}_{(SG)}} = +0.17 \pm 0.00$ and ${\rm {[Eu/Fe]}_{(TO)}} = +0.11\pm 0.00$\\
      for 47~Tuc. 
      We can see that there is almost no difference between the subgiants and the TO stars. Giants in other 
      GCs give very similar results (oversolar values near +0.40--0.50 dex), as seen from 
      \citet{Armos1994}, \citet{Shet1996}, \citet{Sneden1997,Sneden2000,Sneden2004}, 
      \citet{Ivans1999,Ivans2001}, \citet{Shet2000}, \citet{Lee2002} and \citet{Ram2003}. 
      The dispersion in the [Eu/Fe] ratio among all these Galactic globular 
      clusters is extremely small in comparison to the existing scatter of the [Ba/Fe] ratio 
      in the same clusters.
\end{description}

We did not observe any systematic effect between our field halo stars and previous 
analyses of field stars \citep[we used data from][]{Burris2000,Fulb2002,Mash2003} in the same metallicity 
range ($-3.0 \la [{\rm Fe/H}] \la -0.5$). Consequently, as our cluster stars have been observed 
in the same conditions as the field star sample, we did not expect to find any systematic effect in the 
abundance ratios found for the cluster stars.

The choice of analysing subgiants and especially dwarves near the TO point in the three 
clusters is very interesting. As a matter of fact, most of the previous studies of globular 
clusters have been made with bright giant stars ($V \sim 11$--12 mag), which have very strong and 
broad lines. To produce accurate abundances for heavy ($n$--capture) elements in such stars is not easy 
for several elements like strontium (strong and broad lines) or europium (weak and blended lines). With 
our sample of fainter stars ($V \sim 15$--17.5 mag), we have obtained accurate abundances for barium and 
yttrium, but also for the strong lines of strontium and even for europium. 
The abundances of yttrium, barium and europium are in fair agreement with previous abundance determinations 
made in a couple of giants in 47~Tuc, NGC~6752, and NGC~6397 \citep{Norris1995,Cast2000}. To our knowledge, 
these are the first precise measurements of heavy element abundances in such a sample of unevolved stars, 
and these are probably also the first results for strontium abundances in these three GCs.

For the three GCs, the standard deviation around a mean ratio (see Table~\ref{ClusterAbundances}) 
gives a good idea of the internal errors (excitation potentials, $gf$, equivalent widths, line fits...) 
in all the lines used for the abundance determinations. 
When combined with the errors in the atmospheric parameters (see Table~\ref{TabErrors}), it gives an 
estimation of the total error in the abundance determination of a given element. 
The errors in the individual [$n$--capture/Fe] ratios are here usually a bit larger than the 
difference between the ratios in the subgiants and in the TO stars, and if combined with the 
errors in the atmospheric parameters they definitively are larger. For strontium, yttrium 
and europium, this means that the data could be compatible with a unique mean value for each 
cluster (with the exception of the [Y/Fe] ratio in 47~Tuc that shows a difference of 0.17 dex 
between subgiants and TO stars). The case of barium seems a bit different: the [Ba/Fe] ratios 
systematically show small offsets (0.13--0.14 dex) between the subgiants and the TO stars in our three 
clusters, but the origin of this difference (real effect, quality of the spectra, physical models, 
non-LTE effects, details of the hfs correction -which varies strongly with the 
line strength-... ) is still unclear. This difference also seems to be present in M~5, which is one of the 
very few other GCs where abundances were determined simultaneously in a small group of 
subgiants and in stars near the turnoff: \citet{Ram2003} observed 5 subgiants ($V \sim 17$ mag) 
and 6 TO stars ($V \sim 18$ mag). The mean value (from the data in their Table 5D) for the [Ba/Fe] ratio 
in M~5 should be $-0.17 \pm 0.09$ in the subgiants and $-0.33 \pm 0.25$ for the stars near the TO point. 
It would however be very hazardous to reach any general conclusion on this matter because 
\citet{Ram2003} did not present abundances of other $n$--capture elements 
in their sample of subgiants and in the TO stars of M~5, so that no further comparison can 
be done.\\

As a conclusion, we can again mark the excellent agreement between the [$n$--capture/Fe] ratios 
in the three GCs NGC~6397, NGC~6752 and 47~Tuc, and in the field halo stars. 
In addition, the comparison to other Galactic GCs for which [Ba/Fe] and [Eu/Fe] 
ratios were also available showed practicaly no dispersion in the [Eu/Fe] ratios and 
excellent agreement in the [Eu/Fe] ratios between all the GCs at different 
metallicities. This could mean that these Eu abundances were already fixed in the Interstellar 
Medium (ISM) when the clusters formed. An existing dispersion in the [Ba/Fe] ratios similar to that 
of the field halo stars is also seen in this comparison, indicating a possible complex origin of 
the Ba abundances, e.g. involving contributions from early generations of intermediate-mass stars 
in the clusters.

\subsection{Neutron-capture elements ratios}

The $r$--process element Eu is frequently compared to Ba, which is representative of the 
$s$--process in the Sun. Previous analyses of metal-poor stars 
\citep{Gra1994,McW1995,McW1998,Burris2000,Fulb2002} have shown that the [Ba/Eu] ratio decreases at 
low metallicities and approaches a pure $r$--process ratio in most of the very metal-poor stars 
\citep[see also the reviews by][]{Truran2002,Cowan2004}. 
This ratio can thus be used to test the relative importance of the $r$-- and $s$--processes to the 
initial mix of materials at the birth of the stars, being very helpful in identifying the 
$n$--capture nucleosynthesis mechanisms that are responsible for the production of the heavier 
elements ($Z \ge 56$) in field stars but also in GCs.

While the abundance patterns of the heavy $n$--capture elements in field halo stars are commonly explained 
with a $r$--process synthesis at very low metallicities followed by a progressive enrichment due to 
\textit{main s}--process contributions at higher metallicities, the origin of the light ($30 < Z < 56$) 
$n$--capture elements such as Sr or Y, which can be also produced by a \textit{weak s}--process, is more 
complex and is believed to involve at least one additional nucleosynthesis source 
\citep[][ and references therein]{Burris2000,Honda2004,Travaglio2004}. It is therefore interesting to study 
the ratios of the lighter to the heavier $n$--capture elements in order to constrain these formation processes.

\begin{figure}
  \centering
  \includegraphics[width=8.5cm]{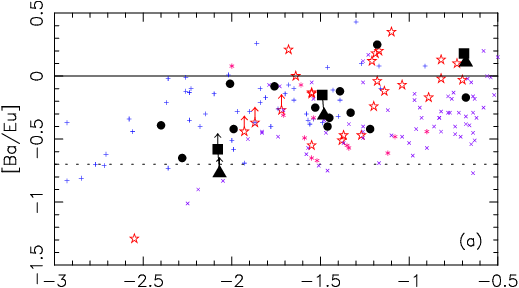}\\ \vspace{0.1cm}
  \includegraphics[width=8.5cm]{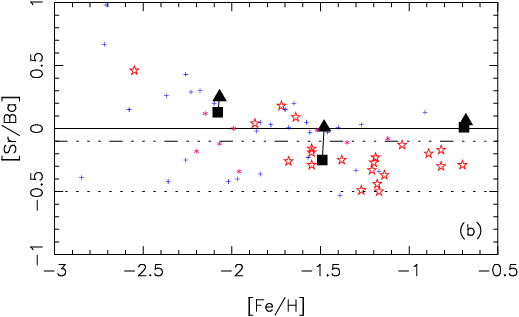}
  \caption{Comparison of the [Ba/Eu] (\textit{Panel a}) and [Sr/Ba] (\textit{Panel b}) 
    ratios for Galactic globular clusters and field halo stars 
    (same symbols than in Figure~\ref{GCvsF_all}).}
  \label{Heavy_Ratios}
\end{figure}

These processes are also likely to happen in GCs; it is therefore also very important to extend such 
analyses to these peculiar objects and to compare their behaviour to that of the field stars with similar stellar 
parameters. Figure~\ref{Heavy_Ratios} shows the [Ba/Eu] and [Sr/Ba] abundance ratios as a function of the 
metallicity for our programme stars (plus additional data from the literature). The horizontal full line in both 
panels (Fig.~\ref{Heavy_Ratios}a and \ref{Heavy_Ratios}b) marks the total solar abundance ratio 
($[{\rm Ba/Eu}] \equiv 0$ and $[{\rm Sr/Ba}] \equiv 0$).

\begin{description}
\item $[{\rm Ba/Eu}]$ \textit{(Fig.~\ref{Heavy_Ratios}a)}: 
      the dotted line marks the solar $r$--process-only fractional ratio 
      ${[{\rm Ba/Eu}]}_{(r-only)} = -0.70$, which is obtained using the data of 
      \citet{Arlandini1999}. 

      A first quick look at this figure reveals a dispersion in the [Ba/Eu] ratio. This is mainly a direct 
      consequence of the previously described dispersion in the Ba abundances in our field stars 
      as well as among the different clusters (see Fig.~\ref{GCvsF_all}c and \ref{GCvsF_all}d). Regarding 
      our sample of field stars, the [Ba/Eu] ratios are distributed near or below the solar value, between the 
      solar \textit{r+s} mix and the solar \textit{r--only} fractional ratio, and are fully compatible 
      with previous analyses in this metallicity range \citep[shown in the figure:][]{Burris2000,Fulb2000,Mash2001,Mash2003}. 

      Our three GCs clearly show a decline of the [Ba/Eu] ratio with decreasing metallicities, 
      which is similar to that observed in the field stars. But we must still be cautious with that assumption. 
      In fact, when we add other GCs for comparison, the dispersion among the clusters makes 
      it more difficult to draw a final conclusion: 47~Tuc (this work) and M~71 \citep{Ram2003}, which 
      are the most metal-rich ($[{\rm Fe/H}] \simeq -0.7$) clusters in this figure, have [Ba/Eu] ratios 
      very close to that of the solar \textit{r+s} mix; GCs at intermediate metallicity 
      ($-2.0 \la [{\rm Fe/H}] \la -1.0$), like NGC~6752 \citepalias[see also][]{James2004}, 
      M~5 \citep{Shet1996,Ivans2001,Ram2003}, M~79 \citep{Francois1991}, NGC~362 \citep{Shet2000}, M~13 
      and M~3 \citep{Sneden2004} show clearly undersolar ratios slightly lower than that of M~71, but 
      still higher than the solar \textit{r--only} fractional ratio, while 
      NGC~6541 \citep{Lee2002} and NGC~288 \citep{Shet2000} have ratios closer to that of the solar mix, 
      and M~4 \citep{Ivans1999} is oversolar (but this dispersion is still compatible with the dispersion among 
      field stars); the most metal-poor clusters ($[{\rm Fe/H}] \la -2.0$), NGC~6397 (this work), NGC~6287, 
      NGC~6293 \citep{Lee2002}, M~92 \citep{Shet1996} and M~15 \citep{Sneden1997} have undersolar ratios equal to 
      or even lower than all the previously cited clusters; in particular NGC~6397 
      and M~92 have [Ba/Eu] ratios very close to the solar \textit{r--only} fractional ratio. Finally, we can 
      add that the dispersion between the dwarves and the subgiants in our three programme clusters is very small 
      and that the data are compatible with a unique mean point for each cluster.
      \\

\item $[{\rm Sr/Ba}]$ \textit{(Fig.~\ref{Heavy_Ratios}b)}: 
      the solar $r$--process-only fractional ratio ${[{\rm Sr/Ba}]}_{(r-only)} = -0.50$ is also 
      represented here by a dotted line, and the dash-dotted line indicates the higher ratio 
      $[{\rm Sr}_{(r+w)}/{\rm Ba}_{(r)}] = -0.10$ that is found when including a 
      \textit{weak} $s$--process contribution to the solar strontium, according to 
      \citet{Arlandini1999}.

      After a first inspection of this figure, we notice that the dispersion of the [Sr/Ba] ratio in the field 
      stars seems lower than the dispersion in Sr or Ba taken separately (see Fig.~\ref{GCvsF_all}a and 
      \ref{GCvsF_all}c). Our sample of field halo stars follows exactly the pattern of previous studies in field 
      stars at these metallicities \citep{Burris2000,Mash2001} and it shows an increase in the [Sr/Ba] ratio at 
      the lowest metallicities. For $[{\rm Fe/H}] \ga -1.5$, the [Sr/Ba] ratio is slightly below the solar system 
      \textit{r+s} mix, and at lower metallicities it increases and becomes oversolar, but the dispersion is also 
      larger. In any case, the [Sr/Ba] is always higher than the solar \textit{r--only} fractional ratio.

      Our three programme clusters also show abundance ratios which are similar to those of field halo stars, 
      and the data also seem to indicate a higher [Sr/Ba] ratio in NGC~6397, which is the most metal-poor cluster 
      in our programme. To our knowledge, there are no other measurements of Sr in GCs that we could 
      have used as a comparison, so that we cannot give any final conclusion on the general behaviour of the [Sr/Ba] 
      ratio in GCs. The dispersion in these clusters is also very small. The difference in the [Sr/Ba] 
      ratio between the dwarves and subgiants of NGC~6752 seems to be mainly due to observational errors 
      and has already been discussed in \citetalias{James2004}.

\end{description}

Both ratios in GCs are compatible with field stars in the same metallicity range, but it is 
also interesting to compare the most metal-poor as well as the most metal-rich GCs with very metal-poor 
and $r$--elements rich stars like CS~31082-001 ($[{\rm Fe/H}] = -2.90$, \citealt{Hill2002}) and CS~22892-052 
($[{\rm Fe/H}] = -3.10$, \citealt{Sneden2003}), which both show $n$--capture ratios very close to a pure $r$--process 
origin. In particular, the GCs M~5, M~79, NGC~6293 and M~15 show [Ba/Eu] ratios very close to that of 
CS~31082-001 ($[{\rm Ba/Eu}] = -0.46$), while two of the most metal-poor clusters, NGC~6397 and M~92, have ratios very 
similar to that of CS~22892-052 ($[{\rm Ba/Eu}] = -0.65$). But none of these clusters present a [Sr/Ba] ratio close 
to those of these two stars ($[{\rm Sr/Ba}] = -0.52$ in CS~31082-001, and --0.39 in CS~22892-052). The most 
metal-rich clusters, 47~Tuc and M~71, show abundance ratios that depart significantly from those two $r$--process 
enriched metal-poor stars.\\

To summarize, the [Ba/Eu] and [Sr/Ba] ratios in GCs follow the trends found in field stars in the 
metallicity range $-2.5 \la [{\rm Fe/H}] \la -0.5$, clearly showing an enrichment process that is incompatible with 
an \textit{r--only} origin of the $n$--capture elements. This implies definitively that $s$--process contributions 
must be taken into account for the synthesis of the heavy elements in GCs. To our knowledge, there 
have been (almost) no other precise determinations of Sr abundances in GCs until now, so that we cannot 
compare all of our data to previous studies. But our results seem to support strongly the idea of having several 
different nucleosynthesis sources for the heavy elements in GCs (at the least, the ``classical'' 
$r$--process, and the {\it main} and/or {\it weak s}--processes).

\section{Discussion}

\subsection{About globular clusters and field stars}

For most of the very metal-poor field halo stars ($[{\rm Fe/H}] \la -2.5$) it is commonly assumed that 
$n$--capture elements have been created exclusively by the $r$--process occurring during Type II supernovae explosions 
(``\textit{r--only} hypothesis''). While the first suggestions of \citet{Truran1981} had only 
been based upon data for a few individual metal-poor stars \citep[e.g.][]{Spite1978}, more recent systematic 
observations of heavy elements in extended samples seem to support strongly this hypothesis 
\citep[e.g.][ and references therein]{Gilroy1988,Ryan1996,Burris2000,Fulb2002,Johnson2002,Francois2003}. 
For such metal-poor stars, it is assumed that the ``classical'' $r$--process is alone responsible for the production 
of the heavier ($Z \ge 56$) $n$--capture elements, while the large and increasing scatter in the [Sr/Ba] ratio 
at very low metallicities, e.g. as seen in the data of \citet{Honda2004}, reveals a possible second type of $r$--process 
for some of the lighter ($30 < Z < 56$) $n$--capture elements (see the recent reviews of \citealt{Truran2002}, and 
\citealt{Cowan2004}). This could be interpreted by different mass ranges of supernovae being responsible 
for the production of the light or the heavy $n$--capture elements \citep{Qian2000,Wanajo2003}. 
At higher metallicities, other events begin to contribute to the production of the heavy elements in the Galactic halo. 
Many neutron-capture elements are then also produced by the {\it main s}--process which occurs during the He-burning 
phases of low- or intermediate-mass AGB stars of typically 1--4 $M_\odot$. 
Several light $n$--capture elements like strontium or yttrium can also partly be produced by an additional 
({\it weak}) $s$--process source, during the He-core burning of massive stars 
\citep[see also][ and references therein]{Busso1999,Travaglio2004}.

Concerning the origin of the heavy elements in GCs in comparison to the field halo stars, many questions 
may arise, but we will concentrate mainly on these: 
what is the origin of their metallicities? Are GCs self-enriched? How can they be so homogeneous in Fe-peak elements 
and neutron-capture elements, but show such high dispersions in the light metal abundances? 
Do heavy elements in globular clusters share a common origin with field halo stars of similar metallicities? 
Until now, no models and no predictions have been made on this subject, leaving these questions still unanswered. 

Let us briefly summarize some of the latest results obtained from the abundance analyses of globular cluster 
scarcely evolved stars: 
\begin{enumerate}
\item Correlations or anticorrelations may exist between several light metals in globular cluster stars (e.g. O--Na, 
      and Mg--Al) at all evolutionary phases \citep{Gra2001,Ram2003,Carr2004}.

\item Neutron-capture elements do not show such relations as a function of other (lighter) elements, and this is also 
      independent of the evolutionary phase \citepalias[ and this work]{James2004}.

\item There is almost no variation in the abundances of heavy elements between stars at different evolutionary 
      phases in a given globular cluster \citepalias[][ and the present work]{James2004}.
\end{enumerate}

The third point seems to be confirmed not only in our three programme clusters, but also in M~71 and M~5, 
at least for barium, which is the only heavy element presented by \citet{Ram2003} in stars fainter than the RGB. 
We expect that this shall be confirmed by future analyses in other clusters. For the present work, it supports 
strongly the idea that we can compare directly our sample of faint turnoff stars and subgiants to bright giants 
in other GCs (see Sect.~\ref{Results}).

The so-called abundance ``anomalies'' of the light metals in GCs can be interpreted in the light 
of self-pollution scenarios \citep{Cottrell1981}, which predict that these inhomogeneities arise from the mass lost 
by intermediate-mass (4--6 $M_\odot$) stars during their AGB phase \citep{Ventura2001,Dantona2002}. 
Some of these models can explain and/or predict very well the existing anticorrelations between light metals, 
but until now none of them clearly explains the origin of the cluster's metallicities, nor do they give any clues 
on the r--/s--process elements ratios. 
Moreover, \citet{Fenner2004} recently developed a self-consistent model of the chemical evolution 
of intracluster gas to test wether the observed ``anomalies'' of NGC~6752 have been caused by contamination 
from intermediate-mass AGBs. Their model does not distinguish the case where cluster stars form from gas already 
contaminated by AGB ejecta, and the case where existing stars accrete AGB material. They conclude that, for their 
particular model, intermediate-mass AGBs do not seem to be the major cause of all the observed globular cluster 
abundance anomalies. 
It is obviously not yet clear if GCs are completely formed out of already enriched matter, 
and if the heavy element content of the cluster has a primordial origin. Additional models are required, 
starting e.g. from the ``classical'' scenario of self-enrichment proposed by \citet{Cayrel1986} or \citet{Truran1991}.

\subsection{The ``classical'' self-enrichment scenario}

For GCs as for field stars, a first generation of massive stars ($M \ga 10\ M_\odot$) of short lifetime 
($\tau \la {10}^8$ years) and the associated Type II SNe are expected to be the major source of nucleosynthesis 
on a timescale compatible with a halo collapse timescale ($< {10}^8$--${10}^9$ years). However, the most metal-poor field 
stars and the GCs could have different origins and could have been independently contaminated by the ejecta 
of massive stars and Type II SNe. But if we assume a common origin, different models of self-enrichment could be adopted 
for the clusters. \citet{Cayrel1986} and \citet{Truran1991} proposed simple models of self-enrichment where 
the metallicity distribution of GCs is a superposition of that of a metal-poor, primary, and possibly 
primordial system, and that of a secondary system born from matter processed by the primary system. In these scenarios, 
the metallicity of the old halo clusters can be understood as caused by self-pollution of primordial ($Z = 0$) clouds: 
the first zero-metal stellar generation is polluted by massive stars ($M \ga 10\ M_\odot$) and Type II SNe formed 
in these massive collapsing clouds. The next generation of stars formed the currently observed generation 
in the halo GCs. The abundance ratios of younger GCs formed during the flattening 
of the protogalaxy is explained along the classical view of progressive metal enrichment. 

\citet{Truran1988} proposed a test to see wether this kind of self-enrichment scenario is appropriate for GCs, 
provided by careful observations of abundance patterns in the most metal-rich clusters associated with the halo 
population. Field stars at comparable metallicities show abundance patterns relative to iron which are evolving toward 
those of solar system matter. This reflects the nucleosynthesis contributions from intermediate-mass stars of longer 
lifetimes that enriched the ISM in C, N, Fe--peak nuclei and $s$--process heavy elements. But self-enrichment 
of a globular cluster should happen on a much shorter timescale. Thus, if self-enrichment indeed occurred for the halo 
GCs, then even the most metal-rich clusters should exhibit both the high [$\alpha$/Fe] ratios 
and the $r$--process dominated heavy element patterns which characterize the ejecta of massive stars, as they are seen 
in the spectra of the most metal-poor field halo stars.

This means that the heavy element ratios [Ba/Eu] and [Sr/Ba] should reflect an $r$--process origin, e.g. like 
the peculiar stars CS~31082-001 \citep{Hill2002} and CS~22892-052 \citep{Sneden2003}. But this pattern is clearly 
not seen among GCs, independently of the metallicity. The [Ba/Eu] ratios indicate clearly a growing contribution 
of the {\it main s}--process to the chemical enrichment of GCs, and  the [Sr/Ba] ratios confirm this 
by departing significantly from a pure $r$--process value (see Figs.~\ref{Heavy_Ratios}a and \ref{Heavy_Ratios}b). 
The fact that some of the most metal-poor clusters show [Ba/Eu] ratios very close to a pure $r$--process value 
but a [Sr/Ba] ratio that departs significantly from the $r$--process only ratio could also indicate an additional 
source for the production of strontium (the {\it weak s}--process, which can show some contributions at these 
metallicities). 
The presence of anticorrelations between several $\alpha$--elements (O--Na, Mg--Al) 
in the clusters is also incompatible with a simple self-enrichment model. A more elaborate model is required 
to explain the formation of the halo GCs. However, the idea that abundance ratios of younger GCs can be explained 
``along the classical view of progressive metal enrichment'' is still interesting. As a matter of fact, 47~Tuc 
is $\sim$2.6 Gyr younger and much more metal-rich than NGC~6752 or NGC~6397 \citep{Gra2003c}, so that it seems 
to indicate that younger (disk) clusters must definitively have been formed out of pre-enriched matter 
in a similar way to field stars at high metallicities.

\subsection{Alternative scenarios}

\citet{Thoul2002} continued the development of the EASE scenario 
({\it Evaporation/Accretion/Self-Enrichment}) of \citet{Parm1999} that explains how accretion could 
have altered the stellar surface composition of GCs and halo metal--poor stars. 
According to the EASE scenario, an important fraction of the metal-poor halo stars was initially formed in GCs, 
where they accreted the $s$--process enriched gas produced by cluster stars of higher-masses. This process 
modified their surface chemical composition. Later these stars evaporated from the cluster and formed 
a part of the current halo population. 
The evolution of GCs is described in two phases: (1) ``first generation'' very metal-poor stars have rather 
high masses ($M \simeq 3$--60 $M_\odot$). The most massive stars quickly evolve to become supernovae and enrich 
the intracluster medium in $\alpha$-- and $r$--process elements. These supernova remnants enrich the primordial 
ISM gas and trigger star formation. If the gravitational potential of the proto-cluster is too weak, the stars 
evaporate and become field halo stars. If the gravity is strong enough, a GC is formed. (2) At this point, 
the chemical composition of the intracluster gas is definitively fixed in $\alpha$-- and $r$--process elements. 
Intermediate-mass stars become AGB stars and produce $s$--process elements, which are brought to their surface 
by the third dredge-up. Stellar winds may eject these elements into the surrounding intracluster medium. 
This $s$--process enriched gas can eventually be accreted by low-mass main sequence stars and pollute their 
surface chemical composition. 

Our results show very homogeneous [Eu/Fe] ratios in all GCs. This could be in agreement at least with 
the first phase of the EASE scenario (first generation of high-mass stars, and fixed composition in $r$--process 
elements). But if this accretion mechanism is really efficient, it is expected to affect especially the star's 
convective zone. When a star evolves and becomes a giant, its convective zone becomes much larger, so that 
a dilution effect is expected in the $s$--process enrichment between main sequence turnoff stars and stars 
at the top of the giant branch. Our results show that the content in $s$--process elements may be very different 
from one cluster to the other, but that it does not vary inside a given cluster. It seems on the contrary that 
GCs are quite homogeneous in their heavy element content ($r$-- and $s$--process elements), independently 
of the evolutionary phase of the star. Our results for turnoff stars and subgiants in 47~Tuc, NGC~6752, 
and NGC~6397 are also compatible with the previous analyses of giants by \citet{Norris1995}, and \citet{Cast2000}. 
We find a very small but systematic dispersion in the [Ba/Fe] ratios between turnoff stars and subgiants in these 
three GCs that we still cannot completely explain (errors in the stellar parameters, in the line parameters, 
or in the fitting method; non-LTE effects -see Sect.~\ref{Results}-). However, if this effect is real, 
it does not support an accretion scenario, as we obtain slightly higher values in dwarves than in subgiants, 
but the dilution in the growing convective zone should produce the opposite results.

In any case, the fact that we find homogeneous abundances of $s$--process elements and simultaneously ``abundance 
anomalies'' in the light metals also in dwarves \citep{Gra2001,Carr2004}, but no correlation between both 
observations, is very interesting because dwarves cannot produce neutron-capture elements alone, nor do they undergo 
deep mixing episodes. To explain both observations at the same time, we could think, for example, that GCs may have 
been formed out of matter which was already enriched in $\alpha$--, $r$-- and $s$--elements, just like ``normal'' field 
halo stars at intermediate or high metallicity. The ``abundance anomalies'' among light 
metals could then be explained with self-pollution mechanisms \citep{Ventura2001,Dantona2002} acting only over 
a restricted mass range 
of the contaminating AGB stars, so that the composition of the GCs in $s$--process elements 
is not altered, or at least only in a very small, nearly undetectable, amount. This could explain why the abundance 
patterns of neutron-capture elements in GCs agree so well with those of field halo stars in the same metallicity range. 
It would also be compatible with the idea that GCs should not always be considered as ``primordial'' objects because, 
after all, we can observe field stars that are up to several hundred times more metal-deficient than the most 
metal-poor GCs \citep{Francois2003,Christ2004}.

\citet{Beasley2003} suggested an alternative scenario in which the most metal-poor halo GCs may have a pregalactic 
origin and may have been enriched by the first supernovae in the universe. They find that one pair-instability 
supernova alone, issued from a very massive primordial star ($140 < M / M_\odot < 260$), could have been sufficient 
to enrich primordial gas to a metallicity of $[{\rm Fe/H}] \sim -2.0$. They do not make predictions for the light metals 
that can be affected by stellar evolution processes (O, Na...), and concentrate only on elements which are heavier 
than silicium. Several problems arise for this kind of model: 
(1) a combination of masses for the Population III stellar progenitors is needed to fit simultaneously the observed 
abundance ratios of [Si/Fe], [Ca/Fe], and [Ti/Fe], [V/Fe], or [Ni/Fe]; 
(2) such massive progenitors cannot produce neutron-capture elements (no $s$-- or $r$--process). 
Additional physics (rotation of the massive star, pollution of the immediate environment) could probably explain 
the presence of neutron-capture elements in GCs, but in any case this scenario hardly explains why the light metals 
also show a huge dispersion in main-sequence turnoff stars...

\section{Conclusions}

We have determined abundances for several neutron-capture elements in stars near the turnoff and at the base 
of the RGB in three globular clusters at different metalicities (NGC~6397, NGC~6752, and 47~Tuc). These data 
have been obtained at the VLT in the framework of the ESO-Large Programme ``Distances, Ages, and Metal Abundances 
in Globular Clusters'', using the high-resolution echelle spectrograph UVES. All the observed stars are fainter 
than $V \sim 15$ in magnitude. These are the first results concerning neutron-capture elements in these scarcely 
evolved stars.

We have used our data to test several scenarios proposed for the formation of globular clusters in the Galaxy. 
From this analysis, we conclude that the ``classical'' self-enrichment scenario as it was proposed 
e.g. by \citet{Truran1991}, cannot explain the observed abundances of neutron-capture elements. 
More elaborate models \citep[e.g.][]{Thoul2002,Beasley2003} still have difficulties in explaining these 
abundance ratios. However, our main results (\citetalias{James2004}, and this work) can be summarized 
as follows:

\begin{enumerate}
\item Globular clusters do not seem to show any clear correlation or anticorrelation in the abundance patterns 
      of the neutron-capture elements or between heavy elements and light metals.

\item Almost all globular clusters are very homogeneous in the abundances of heavy elements, independent 
      of the evolutionary phase of the observed globular cluster stars.

\item The heavy element abundance patterns, and the [Ba/Eu] and [Sr/Ba] ratios, indicate clearly a progressive 
      chemical enrichment in $s$--process elements with increasing metallicity for all globular clusters.
\end{enumerate}

Considering these results, if a self-pollution scenario is assumed to explain the huge dispersions observed 
in the light metals \citep{Ventura2001,Dantona2002}, we could conclude that the contaminating stars should 
not be able to produce sufficient amounts of $s$--process elements to yield a clear, visible, correlation between 
these ``abundance anomalies'' and the neutron-capture elements in globular clusters.

There are still only very few observations of very faint stars in globular clusters, so that many points still 
require stronger supportive evidence. Further observations in other clusters are needed, e.g. to confirm: 
(1) the homogeneity in the abundances of neutron-capture elements at different evolutionary phases, from the main 
sequence to the top of the giant branch; 
(2) if the existing dispersion in the [Ba/Eu] ratio among the most metal-poor ($[{\rm Fe/H}] \la -2.0$) clusters 
is real. 
More modelling is also needed, not only to explain the formation of the clusters, or the strong dispersion observed 
in the light metals, but also including predictions on the neutron-capture elements.

\begin{acknowledgements}
The authors would like to thank the referee, J.~Cohen (California Institute of Technology, USA), 
for her comments. A part of this work was done while G.J. and P.F. were visiting the Sophia University 
and the National Astronomical Observatory in Tokyo, Japan. S.~Wanajo, Y.~Ishimaru, W.~Aoki and S.~Honda 
are warmly thanked for the many very helpful discussions. G.J. would also like to thank R.~Cayrel 
(Obs. de Paris, France) for his advice.

\end{acknowledgements}

\bibliographystyle{aa}


\end{document}